# COSMOLOGY WITH CLUSTERS OF GALAXIES

NETA A. BAHCALL
*Astrophysical Sciences,*
*Princeton, NJ 08544-1001*

Rich clusters of galaxies, the largest virialized systems known, provide a powerful tool for the study of cosmology. Some of the fundamental questions that can be addressed with clusters of galaxies include: how did galaxies and large-scale structure form and evolve? What is the amount, composition and distribution of matter in the universe? I review some of the studies utilizing clusters of galaxies to investigare, among others:
− The dark matter on clusters scale and the mean mass-density of the universe;
− The large-scale structure of the universe;
− The peculiar velocity field on large scales;
− The mass-function of groups and clusters of galaxies;
− The constraints placed on specific cosmological models using the cluster data.

## 1. Introduction

Rich clusters of galaxies can be used effectively to constrain cosmological parameters. I discuss in this paper some of the methods used and their results.
− Optical, X-ray, and gravitational-lensing data of clusters are used to determine the amount of dark-matter in clusters. These data provide the best determination of the dynamical mass-density on ~ Mpc scales. While additional lensing data are needed in order to provide a more accurate comparison of all three methods, the available data suggest $\Omega_{dyn}$ ~ 0.2.
− The mass-to-light ratio of groups and clusters of galaxies is approximately constant for scales $R \gtrsim 0.2 h^{-1}$ Mpc. The flattening of M/L occurs at $M/L_B \simeq$ 200-300h, corresponding to $\Omega$~0.2. The observations suggest that most of the dark matter may be associated with large dark galaxy halos.
− X-ray emission from clusters, which originates from the hot intracluster gas, show that a large fraction of the cluster virial mass is *baryonic* ($\gtrsim$ 30% for h = 1/2, or $\gtrsim$ 10% for h = 1, where $H_o = 100h$ km s$^{-1}$ Mpc$^{-1}$ is the Hubble constant). Combined with the low baryon mass-density determined from nucleosynthesis, this suggests that the total mass-density in the universe may be low, $\Omega$ ~ 0.15 - 0.2.
− The cluster correlation function, an efficient tracer of the large scale structure of the universe, and the cluster mass-function, a tracer of the density fluctuation spectrum, place some of the strongest constraints on cosmological models. For the popular CDM and PBI type models, the density parameter that best fits the cluster observations is, again, $\Omega$ ~ 0.2.

## 2. Clusters of Galaxies: Optical, X-Rays, and Lensing Results

The binding gravitational potential of clusters of galaxies has traditionally been traced by the velocity dispersion of the cluster galaxies using the virial theorem (e.g. Zwicky 1957; Peebles 1980). The cluster potential can also be estimated by the temperature of the hot intracluster gas as determined from X-ray observations (e.g., Jones and Forman 1984; Sarazin 1986) and by the weak gravitational lens distortion of background galaxies caused by the intervening cluster mass (Kaiser and Squires 1993). Sufficient data are now available in both the optical (galaxy velocity dispersion in clusters) and X-ray (the temperature of the intracluster gas) so that these two methods can be compared. In addition, some current lensing data are becoming available. We compare these three independent methods for cluster mass determination.

Optical–X-Rays. The standard hydrostatic, isothermal model for clusters assumes that both the gas and the galaxies in the cluster are isothermal and in hydrostatic equilibrium within a common potential. The relation between the galaxy velocity dispersion ($\sigma$) and the temperature (T) of the intracluster medium (ICM) is then expected to be $\sigma \propto T^{0.5}$. Lubin and Bahcall (1993) analyzed the largest available sample of clusters for which both $\sigma$ and T have been measured. They find a strong correlation between $\sigma$ and T, $\sigma = 332\pm52(kT)^{0.6\pm0.1}$ km s$^{-1}$, where kT is in keV (Fig. 1); the observed relation is consistent with that expected from the isothermal, hydrostatic model. Previous results, obtained with smaller cluster samples (Mushotzky 1984; Edge & Stewart 1991), are generally consistent with these findings.

The observed $\sigma$ (T) relation suggests that the galaxies and the intracluster gas trace the same potential. To determine the average energy per unit mass in the galaxies and the gas, we use the $\beta$ parameter (Mushotzky 1984, 1988; Sarazin 1986; Edge & Stewart 1991), $\beta = \sigma^2/(kT/\mu m_p)$, where $\mu m_p$ is the average particle mass, $\sigma$ is the one-dimensional rms velocity dispersion of the galaxies, and kT is the temperature of the ICM. In the self-gravitating isothermal model (assuming an isotropic velocity distribution), $\beta = 1$; this implies that the galaxies and the ICM trace the same potential and that the energy per unit mass in the gas and in the galaxies are equal. Lubin and Bahcall (1993) used the observed $\sigma$(T) relation to determine the best-fit $\beta$ parameter for clusters; they find $\beta = 0.94\pm0.08$, and a median of $\beta_{(med)} = 0.98$. These values suggest that the galaxies and gas trace each other with little or no velocity bias in the average cluster. It also suggests that both the galaxies and the gas are good tracers of the underlying cluster mass.

The comparable kinetic energies of the gas and galaxies in clusters suggest that the gas and galaxies should also have similar density profiles in the clusters. Bahcall and Lubin (1993) find that this is indeed satisfied thus resolving the long-standing "$\beta$–discrepency" problem in clusters. The average gas and galaxy distributions in clusters are therefore consistent with an approximate hydrostatic, isothermal cluster model where both the galaxy velocities and the gas temperature reflect the underlying cluster potential.

Lensing. Recent observations of weak gravitational distortions of background galaxies by intervening cluster masses (Kaiser and Squires 1993) yield positive detections, thus providing estimates of the lensing cluster mass within a given radius (see Kaiser, this proceeding). Smail *et al*. (1994) used this method to investigate two luminous X-ray clusters of galaxies, CL1455+22 at z=0.26 and CL0016+16 at z=0.55; Falham *et al*. (1994) investigated the X-ray cluster MS1224+20 at z=0.33; and Kaiser (this proceedings) studied the cluster A2163 at z=0.17. Smail *et al.* find that the mass distribution derived from the lensing signal in their two clusters is strikingly similar to that traced by the galaxies and by the hot X-ray gas in the clusters. Their inferred cluster mass profile is compared with the galaxy distribution profile in Fig. 2. These findings agree well with our conclusions that the galaxies, gas, and mass approximately trace each other. They find marginal evidence for a greater central concentration of dark matter with respect to the galaxies, but the uncertainties are large. Smail *et al.* also find that the cluster masses determined from lensing are consistent with the masses determined from the hot gas using the hydrostatic assumption. For CL1455+22 they find $M_{cl}(\leq 0.45\ h_{50}^{-1}) \simeq 3.2 \times 10^{14} M_\odot$ from lensing as compared with $3.6 \times 10^{14} M_\odot$ from X-rays; for CL0016+16 they find $\sim 10 \times 10^{14} M_\odot$ from lensing and $7.5 \times 10^{14} M_\odot$ from X-rays, both within $0.6 h_{50}^{-1}$ (all values are for h=0.5). Falham *et al.* (1994) find a considerably larger mass (by a factor of ~ 3) for MS1224+20 from lensing as compared with the virial mass estimate. For A2163, the opposite is found: lensing yields a small or no signal compared with the virial X-ray mass. The lensing versus the optical or X-ray mass comparisons is summarized below:

- CL0016+16    z = 0.55    $r \leq 0.23 h^{-1}$ Mpc    $M_{lens}/M_{opt/X-ray}$ = $0.9 \pm (\sim 50\%)$
- CL1455+22         0.26    $r \leq 0.30$                                  $1.3 \pm (\sim 50\%)$
- MS1224+20       0.33    $r \leq 0.45$                                  $2.9 \pm (\sim 50\%)$
- A2163                   0.17                                                                   <1

The ratio of the gravitational lensing mass to the optical or X-ray mass within the radius r ranges from <<1 to ~3 for the four clusters studied. What can we learn? First, it is clear that a large cluster sample is needed before reliable conclusions can be reached. Individual clusters are *expected* to vary significantly in their mass ratio estimates due to various effects including cluster elongation (elongation along the line-of-sight will yield $M_{lens}/M_{opt/X-ray}$ > 1 while flattening will yield a ratio < 1), anisotropic velocity orbits, etc. Therefore, a mean of a large sample is needed. In addition, effects such as boundary corrections in the lensing mass estimate can produce a lowered lensing mass, while substructure will overestimate the lensing mass (Narayan 1994). Based on the optical and X-ray results and the preliminary lensing results shown above we assume that cluster masses, *on average*, are properly traced by galaxies and gas. More accurate lensing data should clarify this assumption.

## 3. The Mass, M/L, and Baryon Content of Clusters

<u>Cluster Masses and Mass-to-Light Ratio</u>. Detailed optical and X-ray observations of rich clusters of galaxies yield virial cluster masses that ranges from ~$10^{14}$ to ~ $10^{15}$ $h^{-1}$ $M_\odot$ within 1.5 $h^{-1}$ Mpc radius of the cluster center (§2,5). When normalized by the cluster luminosity, a median value of $M/L_B \simeq 300$ h is observed for rich clusters. This mass-to-light ratio implies a dynamical mass density of $\Omega_{dyn} \sim 0.2$. If, as desired by current theory, the universe has critical density ($\Omega$=1), then most of the mass in the universe *can not* be concentrated in clusters, groups and galaxies; the mass distribution in this case would be strongly biased (i.e., does not follow the light).

A recent analysis of the mass-to-light ratio of galaxies, groups and clusters by Bahcall, Lubin and Dorman (1994) suggests that while the M/L ratio of galaxies increases with scale up to radii of R ~ 0.1-0.2$h^{-1}$ Mpc, due to the large dark-halos around galaxies, this ratio appears to flatten and remain approximately constant for groups and rich clusters, to scales of ~ Mpc. The flattening occurs at $M/L_B \simeq$ 200-300h, corresponding to $\Omega \sim 0.2$. This observation may suggest that most of the dark matter may be associated with the dark halos of galaxies and that clusters do *not* contain a substantial amount of additional dark-matter, other than that associated with (or torn-off from) the galaxy halos. Unless the matter distribution is highly biased (with large amounts of dark matter in the "voids" or on very large scales), this suggests that the mass-density in the universe may be low, $\Omega \sim 0.15$-$0.2$.

Baryons in Clusters. Clusters of galaxies contain a large amount of baryons. Within 1.5 $h^1$ Mpc of a rich cluster, the X-ray emitting gas contributes ~ 10 $h^{-1.5}$% of the cluster virial mass (or ~ 30% for h=1/2) (Briel *et al*. 1992). Visible stars contribute only a small addition to this value. Standard Big-Bang nucleosynthesis limits the mean baryon density of the universe to $\Omega_b \sim 0.015 h^{-2}$. This suggests that the baryon fraction in some rich clusters exceeds that of an $\Omega=1$ universe by a large factor. Detailed hydrodynamic simulations (White *et al*. 1993) suggest that baryons are not preferentially segregated into rich clusters. The above results therefore imply that either the mean density of the universe is considerably smaller than the critical value, or that the baryon density of the universe is much larger than predicted by nucleosynthesis. The observed baryonic mass fraction in rich clusters, combined with the nucleosynthesis limit suggest $\Omega \sim 0.15$-$0.2$; this estimate is consistent with $\Omega_{dyn} \sim 0.2$ determined from clusters.

## 4. The Cluster Correlation Function

The correlation function of clusters is one of the most efficient tools in tracing the large scale structure of the universe. Clusters are correlated more strongly than individual galaxies, by an order of magnitude, and their correlation extends to considerably larger scales (~ $50h^{-1}$ Mpc). The cluster correlation strength increases with richness ($\propto$ luminosity or mass) of the system from single galaxies to the richest clusters (Bahcall and Soneira 1983, Bahcall 1988). The correlation strength also increases with the mean spatial separation of the clusters (Szalay and Schramm 1985, Bahcall and Burgett 1986). This dependence results in a "universal" dimensionless cluster correlation function; the cluster dimensionless correlation scale is constant for all clusters when *normalized* by the mean cluster separation.

The two universal relations of the cluster correlation function are represented by (Bahcall and West 1992): (I) $A_i \propto N_i$, and (II) $A_i \simeq (0.4 d_i)^{1.8}$, where $A_i$ is the amplitude of the cluster correlation function, $N_i$ is the richness of the galaxy cluster i, and $d_i$ is the mean separation of the clusters. (Here $d_i = n_i^{-1/3}$, where $n_i$ is the mean spatial number density of clusters i). The relation can be described as a universal dimensionless correlation function of a scale-invariant nature, with a correlation scale $r_{o,i} \simeq 0.4 d_i$.

The $A_i(d_i)$ relation for groups and clusters of various richnesses is presented in Figure 3. The recent automated cluster surveys of APM (Dalton *et al*. 1992) and EDCC (Nichol *et al*. 1992), shown in Fig. 3, are consistent with the predictions of the above relations. The correlation function of X-ray selected ROSAT cluster of galaxies (Romer *et al*. 1994) is also consistent with the relations presented above. Bahcall and Cen (1994) show that the spatial correlation of a flux-limited sample of X-ray selected clusters will exhibit a correlation scale that is smaller than that of a volume-limited, richness-limited sample of comparable apparent spatial density. The flux-limited sample contains poor groups nearby and only the richest clusters farther away. Using the richness-dependent cluster correlations discussed above, Bahcall and Cen (1994) find excellent agreement with the flux-limited X-ray cluster correlations of Romer *et al* (1994) (see Fig. 4).

## 5. The Mass-Function of Clusters of Galaxies

The mass function (MF) of clusters of galaxies, representing the number density of clusters above a threshold mass M, n(>M), constitutes an important test of theories of structure formation in the universe. The richest, most massive clusters form from rare peaks in the initial mass density fluctuations; poorer clusters and groups form from smaller, more common fluctuations. Comparison of the observed abundance distribution of these clusters as a function of their mass with specific model calculations places strong constraints on the cosmological model and its parameters (§6).

Bahcall and Cen (1993) determined the MF of clusters of galaxies using both optical and X-ray observations of clusters. Their MF is presented in Fig. 5. The function can be described analytically by $n(>M) = 4 \times 10^{-5} (M/M^*)^{-1} \exp(-M/M^*) h^3 \text{ Mpc}^{-3}$, with $M^* = (1.8 \pm 0.3) \times 10^{14} h^{-1} M_\odot$, (where the mass M represents the cluster mass within $1.5h^{-1}$ Mpc radius). The use of the MF in constraining cosmological models is discussed in §6.

## 6. Clusters and Cosmological Models

Two of the fundamental observed properties of clusters of galaxies – the cluster correlation function and the cluster mass function – can be used to place strong constraints on cosmological models and the density parameter $\Omega$ by comparison with model expectations. Bahcall and Cen (1992) contrasted these cluster observations with standard and nonstandard CDM models using large N-body simulations ($400h^{-1}$ box, $10^{7.2}$ particles). They find that none of the standard $\Omega = 1$ CDM models, with any bias parameter, can fit consistently both of these cluster observations. A low-density, low-bias ($\Omega \sim 0.2$–$0.3$, $b \sim 1$–$1.3$) CDM-type model, however, provides a good fit to both sets of observations (see Figs 5-8). This is the first CDM-type model that is consistent with the high amplitude and large extent of the correlation function of the Abell, APM, and EDCC clusters. Such low-density, low-bias models are also consistent with other observables, e.g., the angular correlation function of galaxies (Maddox *et al*. 1990), and, with $\Lambda = 1 - \Omega$, the recent COBE results of the microwave background fluctuations (Smoot *et al*. 1992). The $\Omega$ constraints of this result are model dependent; a mixed hot + cold dark matter, for example, with $\Omega = 1$, is also consistent with these cluster data.

The Cluster Mass Function. The observed cluster mass function is presented in Figure 5 together with the CDM simulations results. The standard CDM model ($\Omega = 1$) is presented for several bias parameters b; only the unbiased case, $b \sim 1$, is consistent with the COBE results. This unbiased model, however, is clearly excluded by the observed mass function: it predicts a much large number of rich clusters than is observed. The larger bias model of $b \sim 2$ (White *et al*. 1987), while providing a more appropriate abundance of rich clusters, is too steep for the observed mass function; it is also incompatible with the COBE results.

The low-density, low bias CDM models, with $\Omega \approx 0.25$ and $b \approx 1$, (with or without $\Lambda$) are consistent with the observed mass function (Fig. 5).

The Cluster Correlation Function. How does the CDM model agree with the observations of cluster correlations? Bahcall and Cen (1992) used the large simulations above to answer this question. The model cluster correlation function was calculated for different cluster mean separations $d$, since the correlation function depends on d (§ 4). The most massive clusters in the simulations that comprise a sample of the required mean separation d is used.

The CDM results for clusters corresponding to the rich Abell clusters ($R \geq 1$) with d = 55 $h^{-1}$ Mpc are presented in Figure 6 together with the Bahcall & Soneira (1983) and Peacock & West (1992) observed correlations. The simulations provide ~ 400 $R \geq 1$ clusters − considerably larger than any previously run models. The results indicate that the standard $\Omega = 1$ CDM models are inconsistent with the observations; they cannot provide either the strong amplitude or the large scales ($\geq 50$ $h^{-1}$ Mpc) to which the cluster correlations are observed. Similar results are found for the APM and EDCC clusters.

The low-density, low bias models are consistent with the observed cluster correlation function. They reproduce both the strong amplitude and the large scale to which the cluster correlations are detected. Such models are the only scale-invariant CDM models that are consistent with both cluster correlations and the cluster mass function. The models are also consistent with the APM and EDCC cluster correlations.

The Universal Dimensionless Cluster Correlation.

The dependence of the observed cluster correlation on d (§ 4) was tested in the simulations (Bahcall and Cen 1992). The results are shown in Figure 7 for the low-density model. The dependence of correlation amplitude on mean separation is clearly seen in the simulations. To compare this result directly with observations, we plot in Figure 8 the dependence of the correlation scale, $r_o$, on d for both the simulations and the observations. The low-density model agrees well with the observations, yielding $r_o \approx 0.4d$, as observed (§ 4). The $\Omega = 1$ model, while also showing an increase of $r_o$ with d, yields considerably smaller correlation scales and a much slower increase of $r_o(d)$ (see Fig. 8).

What causes this $r_o \propto d$ dependence? The dependence, seen both in the observations and in the simulations, is most likely caused by the statistics of rare peak events, suggested by Kaiser (1984) as a way of explaining the strong increase of correlation amplitude from galaxies to rich clusters. The correlation function of rare peaks in a Gaussian distribution increases with their selection threshold. Since more massive clusters correspond to a higher threshold, implying rarer events and thus larger mean separation, the above dependence results. A similar dependence $r_o \propto d$ is also expected in a fractal distribution of galaxies and clusters (e.g., Szalay & Schramm 1985).

## 7. Motions of Galaxy Clusters

Clusters of galaxies can also serve as efficient tracers of the large-scale peculiar velocity field in the universe (Bahcall, Gramann and Cen 1994). Using large-scale cosmological simulations Bahcall *et al.* find that clusters move reasonably fast in all models studied, tracing well the underlying matter velocity field on large scales. The clusters exhibit a Maxwellian distribution of peculiar velocities as expected from Gaussian initial density fluctuations. The cluster 3-D velocity distribution, presented in Fig. 9, typically peaks at v ~ 600 km s$^{-1}$ and extends to high cluster velocities of ~ 2000 km s$^{-1}$. The low-density CDM model exhibits somewhat lower velocities (Fig. 9). Approximately 10% of all model rich clusters (1% for low-density CDM) move with v $\gtrsim$ 10$^3$ km s$^{-1}$. A comparison of model expectation with the available data of cluster velocities is presented in Fig. 10 (Bahcall *et al*. 1994). The cluster velocity data is not sufficiently accurate at present to place constraints on the models; however, improved cluster velocities, expected in the near future, should help constrain the cosmology.

Cen, Bahcall and Gramann (1994) have recently determined the velocity correlation function of clusters in different cosmologies. They find that close cluster pairs, with separations r $\lesssim$ 10h$^{-1}$ Mpc, exhibit strong *attrractive* peculiar velocities; the pairwise velocities depend sensitively on the model. The mean pairwise attractive cluster velocities on 5h$^{-1}$ Mpc scale ranges from ~ 1700 km s$^{-1}$ for $\Omega$ = 1 CDM to ~ 700 km s$^{-1}$ for $\Omega$ = 0.3 CDM (Cen *et al* 1994). The cluster velocity correlation function, presented in Fig. 11, is negative on small scales – indicating large attractive velocities, and is positive on large scales, to ~ 200h$^{-1}$ Mpc –- indicating significant bulk motions in the models. None of the models can reproduce the very large bulk flow of clusters on 150h$^{-1}$ Mpc scale, v $\simeq$ 689±178 km s$^{-1}$, recently reported by Lauer and Postman (1994). The bulk flow expected on this large scale is generally $\lesssim$ 200 km s$^{-1}$ for all the models studied ($\Omega$ = 1 and $\Omega \simeq$ 0.3 CDM, and PBI).

## 8. Summary

Clusters of galaxies provide a powerful tool in constraining cosmological parameters. On cluster scales of ~ Mpc, the dynamical estimates of cluster masses - using optical, X-ray, and some gravitational lensing data – suggest low-densities: $\Omega$ ~ 0.2. A large fraction of the cluster virial mass is baryonic: ~ 10% for h = 1. This suggests that either the baryon density implied by nucleosynthesis, $\Omega_b \simeq$ 0.015h$^{-2}$, is underestimated, or the total mass-density of the universe is low: $\Omega$ ~ 0.15 – 0.2. This $\Omega$ estimate is consistent with the dynamical mass estimate from clusters of galaxies.

Comparing fundamental cluster properties such as the observed mass-function and correlation-function of clusters of galaxies with expectations from various cosmological models also suggests that a low-density ($\Omega$h ~ 0.2), low-bias (b ~ 1) model (with a CDM or PBI-type spectrum) best fits the data. This $\Omega$ constraint, however, is model dependent. The models that best fit the cluster correlation and mass functions are low-density CDM (open or flat) and a mixed cold + hot dark matter (with $\Omega$=1). However, when combined with *all* cluster data, including the constraints on $\Omega$ suggested above, a low-density model ($\Omega$h~0.2; open or flat) with a CDM-type spectrum appears to best fit the overall cluster data.

This work is supported by NSF grant AST93-15368 and NASA grant NGT-51295

Figure 1. Cluster velocity dispersion versus intracluster gas temperature. $N_z$ is the number of measured velocities per cluster (Lubin & Bahcall 1993). The best-fit line is $\sigma=332\pm52$ $(kT)^{0.6\pm0.1}$ km s$^{-1}$.

Figure 2. A comparison of the surface density profile of mass (•) (from lensing) and galaxies (O) in the cluster 14455+22. From Smail *et al*. (1994). (h=1/2 in this plot).

Figure 3. The dependence of the cluster correlation amplitude on mean separation (Bahcall and West 1992).

Figure 4. The cluster correlation function of a volume-limited R≥1 cluster sample (faint solid line) and the relevant flux-limited x-ray cluster correlation for that sample (dark and dashed lines). The X-ray cluster observations (Romer *et al*. 1994; dots) are consistent with expectations (Bahcall and Cen 1994).

Figure 5. The mass-function of clusters of galaxies from observations and simulations (Bahcall and Cen 1992, 1993).

Figure 6. The correlation function of rich (R≥1) Abell clusters, with d=55h$^{-1}$ Mpc, from observations and simulations (Bahcall and Cen 1992).

Figure 7. Dependence of the model cluster correlation function on mean separation d.

Figure 8. Correlation length as a function of cluster separation from observations and simulations (Bahcall and Cen 1992).

Figure 9. (left) The cluster 3-D velocity distribution in four cosmological models (Bahcall, Gramann and Cen 1994).
Figure 10. (right) Comparison of the currently observed 1-D velocity distribution of groups and clusters of galaxies with expectation from an Ω=0.3 CDM model. The dotted line is the model expectation; the dashed line is the model convolved with the observed velocity uncertainties (Bahcall *et al*. 1994).

Figure 11. The velocity correlation function of rich (R≥1) clusters for three models (Cen, Bahcall and Gramann 1994).